\documentclass{ws-ijmpa}

\begin{document}

\markboth{Stefan Scherer} {Manifestly Lorentz-Invariant Baryon Chiral
Perturbation Theory}

%
\catchline{}{}{}{}{}
%

\title{BARYON CHIRAL PERTURBATION THEORY IN MANIFESTLY LORENTZ-INVARIANT FORM}

\author{\footnotesize STEFAN SCHERER}

\address{Institut f\"ur Kernphysik, Johannes Gutenberg-Universit\"at Mainz,
J.~J.~Becher Weg 45, 55099~Mainz, Germany}

\maketitle


\begin{abstract}
   A successful effective field theory program requires besides the most
general effective Lagrangian a perturbative expansion scheme for observables in
terms of a consistent power counting method.
   We discuss a renormalization scheme for manifestly Lorentz-invariant
baryon chiral perturbation theory generating a simple and consistent power
counting for renormalized diagrams.
    The approach may be used in an iterative procedure to renormalize
higher-order loop diagrams and also allows for implementing a consistent power
counting when vector mesons are explicitly included.

\end{abstract}

\section{Introduction}    

   Effective field theory (EFT) has become a powerful tool in the description of the
strong interactions at low energies.
   The central idea is due to Weinberg:\cite{Weinberg:1978kz}
"...  if one writes down the most general possible Lagrangian, including all
terms consistent with assumed symmetry principles, and then calculates matrix
elements with this Lagrangian to any given order of perturbation theory, the
result will simply be the most general possible S--matrix consistent with
analyticity, perturbative unitarity, cluster decomposition and the assumed
symmetry principles."
   The application of these ideas to the interactions among the Goldstone bosons
   of spontaneous chiral symmetry breaking in QCD is referred to as
(mesonic) chiral perturbation theory (ChPT) \cite{Weinberg:1978kz,Gasser:1983yg}
and has been highly successful (see, e.g., Ref.~\refcite{Scherer:pf:2002tk} for a
pedagogical introduction).
   Besides the most general Lagrangian a successful EFT program requires a consistent
power counting scheme to assess the importance of a given (renormalized) diagram.
   In the following we will outline some recent developments in devising
a renormalization scheme leading to a simple and consistent power counting for
the renormalized diagrams of a manifestly Lorentz-invariant approach to baryon
ChPT.

\section{Baryon Chiral Perturbation Theory and Power Counting}

\subsection{Illustration}
  The standard effective Lagrangian relevant to the single-nucleon sector
  consists of the sum of the purely mesonic and $\pi N$ Lagrangians,
respectively,\cite{Gasser:1983yg,Gasser:1987rb}
\begin{displaymath}
{\cal L}_{\rm eff}={\cal L}_{\pi}+{\cal L}_{\pi N}={\cal L}_2+ {\cal L}_4 +\cdots
+{\cal L}_{\pi N}^{(1)}+{\cal L}_{\pi N}^{(2)}+\cdots
\end{displaymath}
which are organized in a derivative and quark-mass expansion.
   The aim is to devise a renormalization procedure generating, after
renormalization, the following power counting:
   a loop integration in $n$ dimensions counts as $q^n$,
pion and fermion propagators count as $q^{-2}$ and $q^{-1}$, respectively,
vertices derived from ${\cal L}_{2k}$ and ${\cal L}_{\pi N}^{(k)}$ count as
$q^{2k}$ and $q^k$, respectively.
   Here, $q$ generically denotes a small expansion parameter such as,
e.g., the pion mass.

\begin{figure}
\centerline{\psfig{file=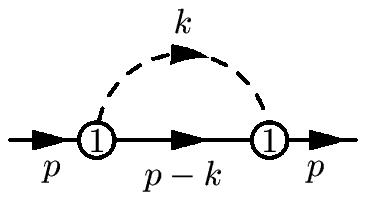,width=4cm}} \vspace*{8pt}
\caption{One-loop contribution to the nucleon self-energy.
The number 1 in the interaction blobs refers to ${\cal L}_{\pi N}^{(1)}$.}
\label{fig:nucleonselfenergypionloop}
\end{figure}

   In order to illustrate the issue of power counting, we consider as an example the one-loop
contribution of Fig.\ \ref{fig:nucleonselfenergypionloop} to the nucleon
self-energy.
   After renormalization, we would like
to have the order
$D=n\cdot 1-2\cdot 1-1+1\cdot 2=n-1.$
   The application of the $\widetilde{\rm MS}$ renormalization scheme of ChPT
\cite{Gasser:1983yg,Gasser:1987rb}---indicated by ``r''---yields
\begin{displaymath}
\Sigma_{\rm loop}^r=-\frac{3 g_{A}^2}{4 F^2}\left[
-\frac{M^2}{16\pi^2}(p\hspace{-.4em}/\hspace{.1em}+m) +\cdots\right]= {\cal
O}(q^2),
\end{displaymath}
where $M^2$ is the lowest-order expression for the squared pion mass.
   The $\widetilde{\rm MS}$-renormalized result does not
produce the desired low-energy behavior which has widely been interpreted as the
absence of a systematic power counting in the relativistic formulation of ChPT.

\subsection{Infrared regularization and extended on-mass-shell scheme}

   Recently, several methods have been suggested to obtain a consistent
power counting in a manifestly Lorentz-invariant approach.
   We will illustrate the ideas in terms of the integral
\begin{displaymath}
H(p^2,m^2;n)= \int \frac{d^n k}{(2\pi)^n} \frac{i}{[(k-p)^2-m^2+i0^+][k^2+i0^+]},
\end{displaymath}
where $\Delta=(p^2-m^2)/m^2={\cal O}(q)$ is a small quantity.
   Applying the dimensional counting analysis of Ref.~\refcite{Gegelia:zz},
the result of the integration is of the form
\begin{displaymath}
H\sim F(n,\Delta)+\Delta^{n-3}G(n,\Delta),
\end{displaymath}
where $F$ and $G$ are hypergeometric functions and are analytic in $\Delta$ for
any $n$.

   In the infrared regularization of Becher and Leutwyler\cite{Becher:1999he} one makes use of
the Feynman parametrization
\begin{displaymath}
{1\over ab}=\int_0^1 {dz\over [az+b(1-z)]^2}
\end{displaymath}
with $a=(k-p)^2-m^2+i0^+$ and $b=k^2+i0^+$.
   The resulting integral over the Feynman parameter $z$ is then rewritten as
\begin{eqnarray*}
I_{N\pi}(-p,0)=\int_0^1 dz \cdots &=& \int_0^\infty dz \cdots
- \int_1^\infty dz \cdots,\\
\end{eqnarray*}
where the first, so-called infrared (singular) integral satisfies the power
counting, while the remainder violates power counting but turns out to be regular
and can thus be absorbed in counterterms.

   The central idea of the extended on-mass-shell (EOMS)
scheme\cite{Gegelia:1999gf,Fuchs:2003qc} consists of performing additional
subtractions beyond the $\widetilde{\rm MS}$ scheme.
   Since the terms violating the power counting are analytic in small
quantities, they can be absorbed by counterterm contributions.
   In the present case, we want the (renormalized) integral to be of the order $D=n-1-2=n-3$.
   To that end one first expands the integrand in
small quantities and subtracts those (integrated) terms whose order is smaller
than suggested by the power counting.
   The corresponding subtraction term reads
\begin{displaymath}
H^{\rm subtr}=\int \frac{d^n k}{(2\pi)^n}\left. \frac{i}{[k^2-2p\cdot k
+i0^+][k^2+i0^+]}\right|_{p^2=m^2}
\end{displaymath}
and the renormalized integral is written as $ H^R=H-H^{\rm subtr}={\cal O}(q) $
as $n\to 4$.

\subsection{Remarks}
\begin{itemlist}
\item Using a suitable renormalization condition one
obtains a consistent power counting in manifestly Lorentz-invariant baryon chiral
perturbation theory including, e.g., vector mesons\cite{Fuchs:2003sh} or the
$\Delta(1232)$ resonance\cite{Hacker:2005fh} as explicit degrees of freedom.
\item  We have formulated the infrared regularization of Becher and Leutwyler\cite{Becher:1999he}
in a form analogous to the EOMS renormalization.\cite{Schindler:2003xv}
\item Using a toy model we have explicitly demonstrated the application of both
infrared and extended on-mass-shell renormalization schemes to multiloop diagrams
by considering as an example a two-loop self-energy
diagram.\cite{Schindler:2003je}
   In both cases the renormalized diagrams satisfy a
straightforward power counting.
\end{itemlist}
\section{Applications}
\begin{figure}
\centerline{\psfig{file=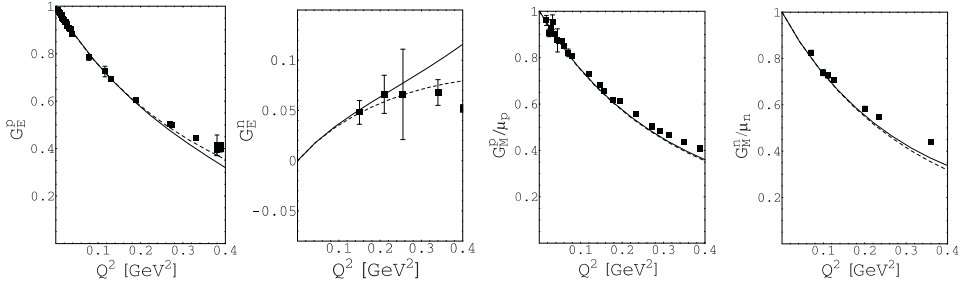,width=\textwidth}} \vspace*{8pt}
\caption{\label{G_neu} The Sachs form factors of the nucleon in
manifestly Lorentz-invariant chiral perturbation theory at ${\cal O}(q^4)$
including vector mesons as explicit degrees of freedom. Full lines: results in
the extended on-mass-shell scheme; dashed lines: results in infrared
regularization.}
\end{figure}
   The EOMS scheme has been applied in several calculations such
as the chiral expansion of the nucleon mass, the pion-nucleon sigma term, and the
scalar form factor,\cite{Fuchs:2003kq} the masses of the ground-state baryon
octet,\cite{Lehnhart:2004vi} and the nucleon electromagnetic form
factors.\cite{Fuchs:2003ir,Schindler:2005ke}
   Figure \ref{G_neu} shows the Sachs form factors of the nucleon in
manifestly Lorentz-invariant chiral perturbation theory at ${\cal O}(q^4)$
including vector mesons as explicit degrees of freedom.\cite{Schindler:2005ke}
   One of the most recent applications has been the derivation of consistency
conditions among the renormalized parameters of the most general EFT
Lagrangian.\cite{Djukanovic:2004mm}
   For example, requiring the consistency of effective field theory with respect to
renormalization results in constraints such as, e.g., the universal $\rho$
coupling.
   Moreover, similar constraints for the QED of vector mesons
predict the gyromagnetic ratio of the $\rho^+$ and the mass difference
$M_{\rho^0}-M_{\rho^\pm}\sim 1$ MeV at tree order.

\section*{Acknowledgments}
I would like to thank D.~Djukanovic, T.~Fuchs, J.~Gegelia, C.~Hacker,
G.~Japaridze, B.~C.~Lehnhart, M.~R.~Schindler, and N.~Wies for their continuous
enthusiasm.

\end{document}